\documentclass[aps,prb,twocolumn, showpacs]{revtex4}

\usepackage{graphicx}
\usepackage{dcolumn}
\usepackage{amsmath}

\begin{document}

\title{On the applicability of bosonization and the Anderson-Yuval
       methods at the strong-coupling limit of quantum impurity
       problems}
\author{L. Borda,$^{1,2}$ A. Schiller,$^3$ and A. Zawadowski$^2$} 
\affiliation{
             $^1$ Physikalisches Institut, Universit\"at Bonn,
                  Nussallee 12 D-53115 Bonn, Germany\\ 
             $^2$ Research Group ``Physics of Condensed Matter''
                  of the Hungarian Academy of Sciences,
                  Institut of Physics,
                  Budapest University of Technology and Economics,
                  Budapest, H-1521, Hungary\\ 
             $^3$ Racah Institute of Physics, The Hebrew University,
                  Jerusalem 91904, Israel} 
\date{\today}
\begin{abstract}
The applicability of bosonization and the Anderson-Yuval
(AY) approach at strong coupling is investigated by
considering two generic impurity models: the interacting
resonant-level model and the anisotropic Kondo
model. The two methods differ in the renormalization
of the conduction-electron density of states (DoS)
near the impurity site. Reduction of the DoS, absent
in bosonization but accounted for in the AY approach,
is shown to be vital in some models yet superfluous
in others. The criterion being the stability of the
strong-coupling fixed point. Renormalization of the DoS
is essential for an unstable fixed point, but superfluous
when a decoupled entity with local dynamics is formed.
This rule can be used to boost the accuracy of both
methods at strong coupling.
\end{abstract}
\pacs{
72.10.Fk,   
72.15.Qm,   
73.63.Kv    
}
\maketitle

---{\em Introduction}.
Several classic models in condensed-matter physics
show logarithmic behavior at high energies, followed
by qualitatively different behavior at low energies.
Notable examples include the x-ray absorption
problem,~\cite{NozieresDeDominicis69} the Kondo
Hamiltonian,~\cite{Kondo64} the interacting resonant-level
model (IRLM),~\cite{VigmanFinkelstein78,Schlottmann82}
and different variants of two-level systems
(TLS).~\cite{YuAnderson84,VladarZimanyi88} Historically
devised to model real impurities in bulk samples,
many of these Hamiltonians have recently found new
realizations and generalizations in quantum dots and
other confined nanostructures.

A distinguished place in the theory of such quantum
impurities is reserved to Abelian
bosonization~\cite{bosonization} and the Anderson-Yuval
(AY) approach,~\cite{AndersonYuval70,AndersonHamann70}
which remain among the most powerful and versatile
analytical tools in this realm. With numerous
applications over the last forty years, it is surprising
that the applicability of neither approach has ever
been studied systematically for strong couplings. In
bosonization, the bare couplings are generally assumed
to be weak. Strong static interactions are often included
{\em ad hoc} in terms of their scattering phase shift.
The AY method, which maps the original impurity
problem onto an effective Coulomb gas, is presumably
nonperturbative in certain couplings. However, it
typically fails to reproduce the correct scaling
equations even at the next-to-leading
order.~\cite{FowlerZawadowski71,Schlottmann82} A
reliable extension of these approaches to strong
couplings is highly desirable.

The goal of the present paper is to critically test the
accuracy of these leading analytic methods away from
weak coupling, and to propose an operational extension
to strong couplings. To this end, we resort to Wilson's
numerical renormalization group~\cite{Wilson75} (NRG), and
to two generic classes of models as test beds: the IRLM
and the anisotropic Kondo model. Our analysis highlights
the role of the reduction in the conduction-electron
density of states (DoS) near the impurity site, which
may hinder the efficiency of other essential couplings
(e.g., tunneling in the IRLM). This reduction of the DoS,
absent in bosonization but included in the AY approach,
proves vital in some models and superfluous in others. It
is essential in cases where the strong-coupling fixed
point is unstable, but superfluous in models where a
decoupled entity with local dynamics is formed at
strong coupling. Hence, the accuracy of bosonization
and the AY approach can be significantly enhanced by
selectively incorporating the DoS renormalization
factor to match the case in question.

The reduction of the local conduction-electron DoS is
best seen for a simple model where electrons scatter
elastically off a point-like impurity ($s$-wave
scattering). The renormalized DoS at the impurity site
takes the form~\cite{MezeiZawadowski71}
\begin{equation}
\varrho(\omega \approx E_{\rm F}) =
        \varrho_0 \cos^2\delta \; ,
\label{eq:dos_change}
\end{equation} 
where $\varrho_0$ is the unperturbed DoS,
$E_{\rm F}$ is the Fermi energy,
and $\delta$ is the scattering phase shift. Since
$\delta \to \pi/2$ for resonant scattering, this
implies $\varrho(\omega \approx E_{\rm F}) \to 0$.
This fact may have a dramatic
effect, as exemplified below by the two-channel IRLM. A
strong local Coulomb repulsion suppresses the DoS at
the vicinity of the impurity, reducing the hopping rate
between the impurity and the bands. Since reduction
of the DoS is independent of the interaction sign, it
equally applies to an alternating potential. The case
of a TLS
with a single coupling (the commutative
model~\cite{YuAnderson84, VladarZimanyi88}) is
qualitatively similar. 

One may expect the same to occur in the anisotropic Kondo
model or the non-commutative TLS with electron-assisted
hopping. For example, consider the single-channel
Kondo model (1CKM) with a large $XXZ$ anisotropy:
$J_z \gg |J_{\perp}|$, with $J_x = J_y = J_{\perp}$. In
the spirit of the AY philosophy,~\cite{AndersonYuval70}
one may first treat the larger coupling $J_z$ before
incorporating the smaller $J_{\perp}$. In the absence of
$J_{\perp}$, a large $J_z$ reduces the local DoS at the
impurity site independent of the orientation of the
impurity spin. Incorporating $J_{\perp}$ at the next
step, its efficiency is expected to be hindered by the
reduced DoS, to the extent that it diminishes in the
limit $J_z \to \infty$ [when $\delta \to \pi/2$ and
$\varrho(\omega \approx E_{\rm F}) \to 0$]. Surprisingly,
this is not what we find with the NRG. Rather, spin
flips remain governed at large $J_z$ by the bare
transverse coupling $J_{\perp}$.

To unravel the governing rule, we conduct a detailed
comparison between Wilson's NRG, bosonization, and
the AY method, applied separately to the multichannel
Kondo and IRLM models. Applicability of the latter
two approaches at strong coupling is shown to depend
crucially on the stability of the strong-coupling limit.
Whenever a decoupled entity with local dynamics is
formed (i.e., a stable strong-coupling fixed point
is reached), then the DoS renormalization factor is
superfluous and bosonization works well. If, however,
the strong-coupling limit is unstable, then the DoS
renormalization factor is essential and the AY
approach works well. The above classification pertains
to non-commutative models. For commutative couplings
the AY method always applies as one can always
reorder the perturbation series.

Prompted by these
findings we proceed to re-examine the ``intimate
relation'' between the IRLM and the anisotropic
1CKM.~\cite{Schlottmann82} Close correspondence
is established between the models in case of the
single-channel IRLM, but not in the case of
multiple screening channels.

---{\em Interacting resonant-level model}.
In the IRLM,~\cite{VigmanFinkelstein78,Schlottmann82}
a 1D electron gas is coupled to a spinless impurity
level by two distinct mechanisms: a hopping matrix
element $V$ and a short-range Coulomb repulsion
$U$. The hopping rate is enhanced for weak repulsion,
but is generally suppressed at large $U$ due to
a reduction in the conduction-electron overlap
integrals between a vacant level and an occupied
one~\cite{GiamarchiNozieres93,BordaZawadowski07} (the
so-called orthogonality catastrophe~\cite{Anderson67}).
Consequently, the hopping rate tends to develop a
maximum at some intermediate coupling $U$, whose value is
pushed toward weak coupling as the number of screening
bands $N$ is increased.~\cite{BordaZawadowski07}
This behavior stems from an enhancement of the
orthogonality effect with increasing $N$.

Interest in the IRLM has been recently rekindled by a
Bethe Ansatz solution of a two-lead version of the
model under nonequilibrium
conditions.~\cite{MehtaAndrei06} In its multichannel
form, the Hamiltonian reads
${\cal H} = {\cal H}_0 + {\cal H}_1 + {\cal H}_2$, with
\begin{eqnarray}
{\cal H}_0 &=& \sum_{n = 0}^{N - 1}
               \sum_{0 < k < 2k_F} v_F (k - k_F)
                     a^{\dagger}_{k n}
                     a^{\phantom{\dagger}}_{k n}
             + \epsilon_d d^{\dagger} d \;,
\label{eq:H_0}\\
{\cal H}_1 &=& U
                 \sum_{n = 0}^{N-1}
                 \bigg(
                        a^{\dagger}_{n}
                        a^{\phantom{\dagger}}_{n}
                        - \frac{1}{2}
                 \bigg)
                 \bigg(
                        d^{\dagger} d - \frac{1}{2}
                 \bigg)\;,
\label{eq:H_1}\\
{\cal H}_2 &=& V
               \left (
                        d^{\dagger} a^{\phantom{\dagger}}_0
                        + a^{{\dagger}}_0 d
               \right )\;.
\label{eq:H_2}
\end{eqnarray}
Here, $a^\dagger_{k n}$ creates an electron with
momentum $k$ in the $n$th band, $d^{\dagger}$
creates an electron on the level, $k_F$ and $v_F$ are
the Fermi momentum and Fermi velocity, respectively,
$\epsilon_d$ is the level energy, $U$ is the Coulomb
repulsion, and $V$ is the tunneling amplitude into
the $n = 0$ band. The operator $a^{\dagger}_{n} =
(1/\sqrt{\cal N}) \sum_k a^{\dagger}_{k n}$, where
${\cal N}$ is the number of lattice sites, creates
a localized band electron at the impurity site. Note
that ${\cal H}$ is particle-hole symmetric for
$\epsilon_d = 0$, the case of interest here.


\begin{figure}
\includegraphics[width=1.0\columnwidth,clip]
                {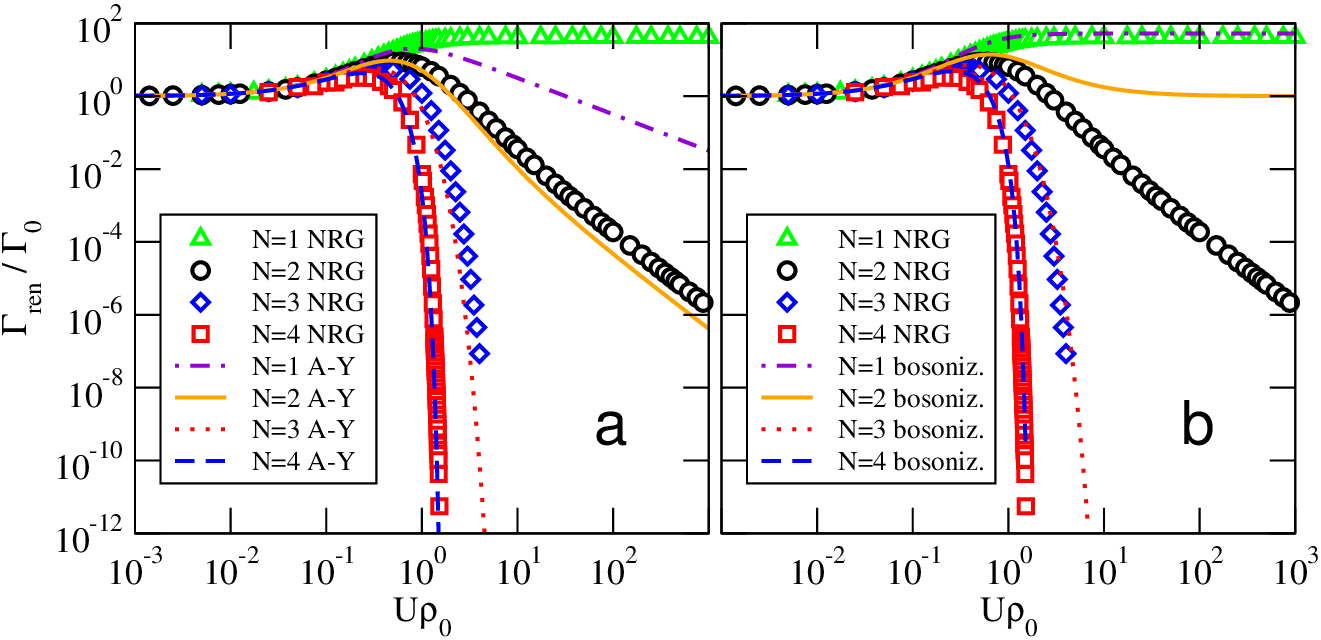}
\caption{(Color online) Renormalized level width for the
         IRLM with up to $4$ screening channels, as obtained
         by the NRG, bosonization, and the AY approach. Here
         $\Gamma_0 = \pi \varrho_0 V^2$ with $V/D_0 = 0.02$
         (we use $\varrho_0 D_0 = 1/2$).
         The AY approach (panel a) works quite well for
         $N = 2, 3, 4$, but fails for $N = 1$. Bosonization
         (panel b) works well for $N = 1, 3, 4$, but
         predicts an exact mapping~\cite{SchillerAndrei07}
         between $U \to 0$ and $U \to \infty$ for $N = 2$,
         and thus a saturated width. Note that the
         AY approach systematically underestimates
         $\Gamma_{\rm ren}$ at large $U$ whereas the
         opposite is true of bosonization.}
\label{fig:resonant}
\end{figure}

We study the IRLM using Wilson's NRG, bosonization
and the AY approach. Since bosonization and the
NRG are frequently used, we refer the reader to
Refs.~\onlinecite{bosonization} and \onlinecite{Wilson75}
for details of these methods. In the following we
briefly review the AY approach, which relies on a mapping
of the impurity problem onto an effective 1D Coulomb gas
of multicomponent charges. The AY mapping is nonperturbative
in the Coulomb repulsion $U$, which determines the
different charge components through its associated
phase shift $\delta = -\arctan(\pi \varrho_0 U/2)$.
Here $\varrho_0$ is the bare conduction-electron
DoS. The hopping amplitude $V$ fixes the fugacity
of the gas, which is given in turn by
\begin{equation}
y = V (\varrho_0\tau_0)^{1/2} \cos\delta .
\label{eq:fugacity}
\end{equation}
Here $\tau_0 = 1/D_0$ is a short-time cutoff, with
$D_0$ the bare bandwidth. The $\cos\delta$ that
appears in Eq.\eqref{eq:fugacity} encodes the DoS
renormalization. A similar mapping, only without the
$\cos\delta$, can be derived using Abelian bosonization.
Incorporating $U$ by means of its associated phase
shift,~\cite{SchillerAndrei07} an identical 1D gas
is obtained with $y = V (\varrho_0\tau_0)^{1/2}$.

The Coulomb gas is next treated by progressively
increasing the short-time cutoff while simultaneously
renormalizing the gas parameters so as to leave
the partition function invariant. This results in
renormalization-group (RG) equations for the parameters
of the Coulomb gas~\cite{BordaZawadowski07} which are
perturbative in the fugacity $y$ (namely, $V$) but
nonperturbative in $U$. To illustrate the basic iterative
step, suppose that the short-time cutoff has already
been increased from its bare value $\tau_0 = 1/D_0$
to $\tau > \tau_0$. Further increasing the cutoff to
$\tau + \delta\tau$ requires two operations:
(i) integration over charge pairs whose separation falls
    in the interval $(\tau, \tau + \delta\tau)$, and
(ii) rescaling of $\tau$ by $\tau + \delta\tau$.
Consecutive charges, having opposite signs, leave no net
charge behind. However, they do possess a dipole moment
that acts to screen the interaction between the charges
that remain. Integration over the close-by charge pairs
can therefore be absorbed into a renormalization of the
remaining charges. On the other hand, the rescaling
of $\tau$ is absorbed into a renormalization of
the fugacity $y$, as described by the following
set of RG equations:~\cite{BordaZawadowski07}
\begin{eqnarray}
\frac{dy}{d\ln\tau} &=& y
         \left(
                \frac{1}{2}
                - z_0 - \frac{1}{2} \sum_{n = 0}^{N-1}
                                          z_{n}^2
         \right) \;,
\label{eq:y_scaling}\\
\frac{dz_{n}}{d\ln\tau} &=& 2
         \left(
                \delta_{n 0} + z_{n}
         \right)y^2 \;. 
\label{eq:z_scaling}
\end{eqnarray}
Here $\delta_{n 0}$ is the Kronecker delta, while
the charge components $z_n$ take the bare value
$z = 2\delta/\pi$.
Contrary to usual dynamical scaling equations,
the DoS is also modified in this procedure
due to the rescaling of $\tau$.
However, this difference is only formal. Either
strategy can be pursued.

Equation \eqref{eq:y_scaling} pertains to the fugacity
$y$. It can equally be written as a scaling equation
for the level width $\Gamma = \pi y^2/\tau$, which
serves as the low-energy cutoff in the problem.
Specifically, the perturbative RG procedure terminates
at $1/\tau \sim \Gamma$, when the fugacity $y$ becomes
of order $1$. Whether this condition is met or not
depends on the values of $N$ and $\delta$. To see
this, consider a sufficiently small $y_0$ such
that the renormalizations of $z_n$ can be ignored.
Equation \eqref{eq:y_scaling} then becomes
\begin{equation}
\frac{dy}{d\ln\tau} =
         \frac{1}{2}
         \left(
                1 - 2z - N z^2
         \right) y .
\label{simplified-RG}
\end{equation}
Whether $y$ is relevant or not depends on the
sign of the expression in the brackets. Since
$-1 < z < 0 $ for repulsive interactions, $y$ is
always relevant for $N \leq 3$. However, it turns
irrelevant for $N > 3$ if $U$ is made sufficiently
large. The system flows then to a decoupled level.
Careful analysis of the transition between a strongly
coupled and a decoupled level shows that it is of the
Kosterlitz-Thouless type,~\cite{SBZA08} analogous to
the ferromagnetic-antiferromagnetic transition line
of the anisotropic Kondo model.
Importantly, bosonization and the AY approach
predict the same critical coupling $U_{\rm c}$ as
$V \to 0$.

\begin{figure}
\includegraphics[width=1.0\columnwidth,clip]
                {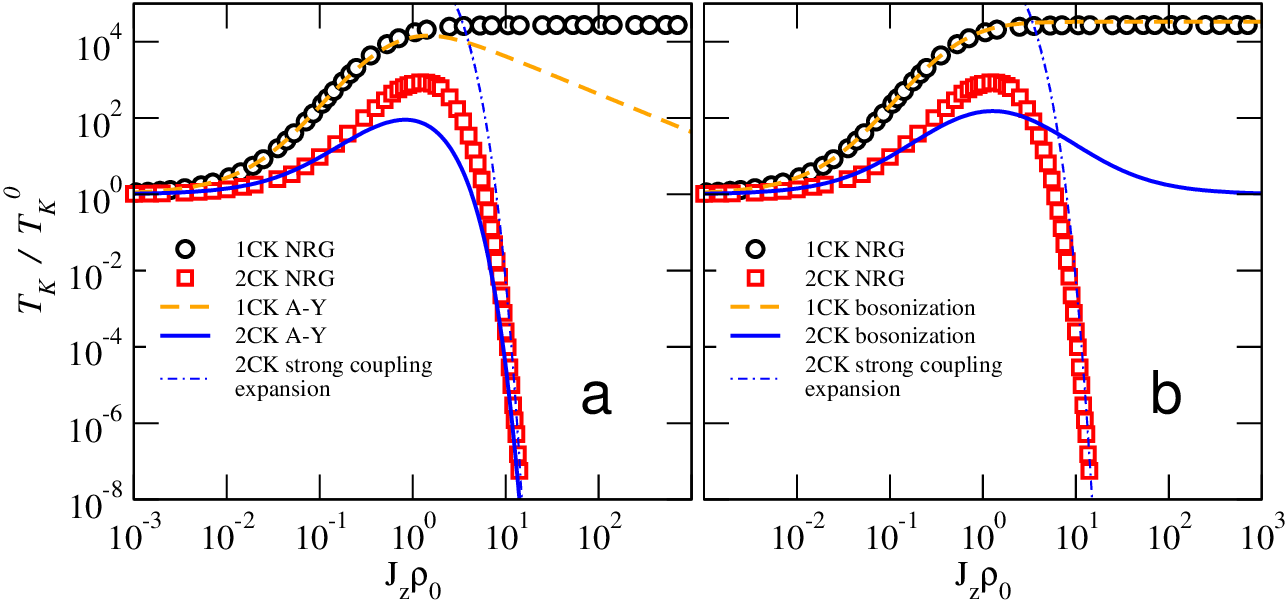}
\caption{(Color online) The Kondo temperature of the one-
         and two-channel Kondo model as a function of
         $J_z$ for fixed $\varrho_0 J_\perp = 0.1$. While
         bosonization works quite well for the 1CKM, the
         AY approach incorrectly predicts a vanishing
         $T_K$ as $J_z \to \infty$. The roles are
         reversed for the 2CKM. Here the AY method is
         qualitatively correct, whereas bosonization
         predicts~\cite{SchillerDeLeo08} an exact mapping
         between $J_z \to 0$ and $J_z \to \infty$, and
         thus a saturated $T_K$.
         The dotted-dashed line shows a one-parameter
         fit [the prefactor of $\Lambda_{\perp}$ in
         Eq.(47) of Ref.~\onlinecite{SchillerDeLeo08}]
         to the strong-coupling expansion of the 2CKM.}
\label{fig:kondo}
\end{figure}

Solution of Eq.\eqref{simplified-RG} in the regime
where $y$ is relevant yields the renormalized level
width, or cutoff scale,
\begin{equation}
\Gamma_{\rm ren} \sim D_0 y_0^{2/(1 - 2z - Nz^2)} .
\label{Gamma-ren}
\end{equation}
Here $y_0$ is the bare fugacity of
Eq.\eqref{eq:fugacity}. For either $N = 1$ or $N = 2$,
one can substitute $z \simeq -1$ in Eq.\eqref{Gamma-ren}
to obtain $\Gamma_{\rm ren} \sim D_0 y_0^{2/(3 - N)}$
at large $\varrho_0 U$. Hence $\Gamma_{\rm ren}$ is
strongly suppressed as $\varrho_0 U \to \infty$
due to the $\cos\delta$ that appears in $y_0$.
By contrast, $\Gamma_{\rm ren}$ saturates in
bosonization, where the DoS renormalization factor
is absent.

Figure~\ref{fig:resonant} compares the renormalized
level width $\Gamma_{\rm ren}$ of the multichannel
IRLM, as obtained by our three methods of interest.
Within the NRG, $\Gamma_{\rm ren}$ was defined from
the $T \to 0$ charge susceptibility of the level according to
$\Gamma_{\rm ren} = 1/\pi \chi_{\rm c}$. In the
AY approach and bosonization, $\Gamma_{\rm ren}$
was obtained from a full numerical solution of
Eqs.\eqref{eq:y_scaling} and \eqref{eq:z_scaling},
with and without the $\cos \delta$ in 
Eq.\eqref{eq:fugacity}.

While both the AY method and bosonization work
quite well for $N > 2$, only the former approach
succeeds in tracing the NRG for $N = 2$. Bosonization
fails to produce the suppression in $\Gamma_{\rm ren}$
at large $U$, which stems from the renormalized DoS.
By contrast, the AY approach fails to generate the
saturation in $\Gamma_{\rm ren}$ for $N = 1$ and
large $U$, which bosonization captures quite well.
Hence, the DoS renormalization factor is superfluous in
this case.  The source of distinction between $N = 1$
and $N = 2$ is nicely elucidated by a strong-coupling
expansion~\cite{SchillerAndrei07} in $1/U$. Whereas a
decoupled entity with local dynamics is formed when
$N = 1$, for $N = 2$ the strong-coupling fixed
point is unstable. A renormalized IRLM is
recovered,~\cite{SchillerAndrei07} with dynamics that
depends on the renormalized DoS. Relevance of the
DoS renormalization depends then on the stability of
the strong-coupling fixed point. As shown below, the
same criterion applies to the Kondo model.

---{\em Anisotropic single-channel Kondo model}.
The anisotropic 1CKM has been intensely studied over
the years~\cite{AndersonYuval70,FowlerZawadowski71,
AndersonHamann70,Solyom74} as a paradigmatic example
for strong correlations. It describes the spin-exchange
interaction of an impurity spin $\vec{S}$ with the
local conduction-electron spin-density $\vec{s}$, as
modeled by the Hamiltonian term
\begin{equation}
{\cal H}_{\rm int} = 
          J_z S^z s^z
          + \frac{J_\perp}{2} \left (
                                      S^-s^+ + S^+s^-
                              \right ) .
\label{eq:H_anis}
\end{equation}
In the antiferromagnetic regime, $J_z > -|J_{\perp}|$,
the system flows to the strong-coupling fixed point of
the isotropic model regardless how large the anisotropy
is. 

Similar to the hopping $V$ in the IRLM, the transverse
Kondo coupling $J_{\perp}$ is attached a factor of
$\cos^2 \delta$ with
$\delta = -\arctan(\pi \varrho_0 J_z/4)$ upon mapping
the 1CKM onto an effective 1D Coulomb gas using the
AY approach. This factor, which stems from the form
of the electronic Green function,~\cite{VladarZimanyi88}
is absent in bosonization, and is omitted in the
original works of Anderson and
collaborators.~\cite{AndersonYuval70,AndersonHamann70}
Its inclusion has profound implications, as the effect
of spin flips (and consequently the Kondo temperature)
vanishes in the limit $\delta \to \frac{\pi}{2}$ (i.e.,
$J_z \to\infty$). If these considerations are
correct, then the NRG should give the same result
as $J_z \to \infty$, which turns out not to be the
case.

Figure~\ref{fig:kondo} compares the Kondo temperature
$T_K$ obtained by our three methods of interest.
Within the NRG, $T_K$ was defined from the $T \to 0$
impurity spin susceptibility according to
$T_K = 1/4 \chi_s$. In the AY approach and bosonization,
it followed from a full numerical solution of the RG
equations~\cite{AndersonHamann70} with and without
the $\cos^2 \delta$ factor attached to $J_{\perp}$.
Evidently, bosonization works
quite well for the 1CKM, reproducing the saturation
of the Kondo temperature as $J_z \to \infty$. The AY
prediction of a vanishing $T_K$ is clearly discredited
by the NRG, proving the redundancy of the DoS
renormalization factor. As anticipated, a decoupled
entity is formed at large $J_z$, signaling the
stability of the strong-coupling fixed point.

A critical test of our picture is provided by the
anisotropic two-channel Kondo model (2CKM), whose
strong-coupling fixed point is known to be unstable.
Instead, the model flows to an intermediate-coupling,
non-Fermi-liquid fixed point, characterized by
anomalous thermodynamic and dynamic
properties.~\cite{CoxZawadowski98} Similar to the
two-channel IRLM, we expect the DoS renormalization
to be essential in this case. The results shown in
Fig.\ref{fig:kondo} well support our picture. While
bosonization predicts~\cite{SchillerDeLeo08} an
exact mapping between $J_z \to 0$ and
$J_z \to \infty$, and thus a saturated $T_K$, the
AY approach correctly reproduces the vanishing of
$T_K$. Though quantitatively less accurate at
intermediate $J_z$, agreement with the NRG is
clearly very good both at small and large coupling.

Above two screening channels, the anisotropic Kondo
model undergoes a Kosterlitz-Thouless transition
with increasing $J_z > 0$ to a ferromagnetic-like
state.~\cite{SchillerDeLeo08} Since spin flips are
suppressed to zero, the distinction between bosonization
and the AY approach looses its significance at strong
coupling, similar to the IRLM with $N > 3$.

\begin{figure}
\includegraphics[width=1.0\columnwidth,clip]
                {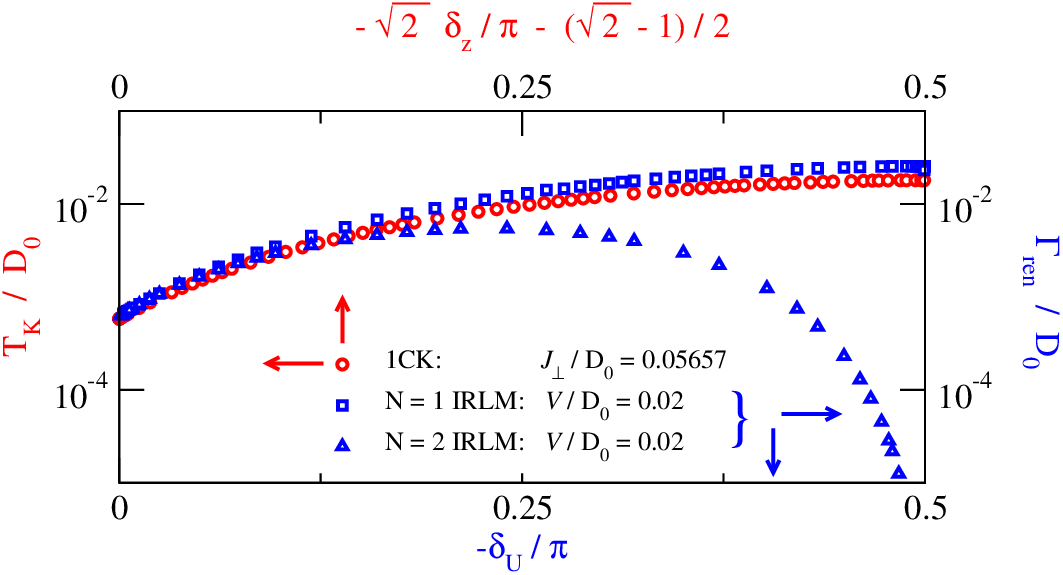}
\caption{(Color online) Renormalized level width of
         the one- and two-channel IRLM vs the Kondo
         temperature of the 1CKM, obtained using the NRG.
         Here $J_{\perp} = \sqrt{8} V$. As predicted by
         bosonization, there is close correspondence
         between the 1CKM and the $N = 1$ IRLM
         upon equating
         $\sqrt{2} \delta_z + \pi(\sqrt{2} - 1)/2$
         with $\delta_{\rm U}$.}
\label{fig:resonant_vs_kondo}
\end{figure}

---{\em Comparison of the two models}.
Prompted by these results, we have set out to carefully
test the accepted mapping~\cite{Schlottmann82} of the
one-channel IRLM onto the 1CKM, as the mapping involves
large couplings. Within bosonization, one finds the
following correspondence of parameters:~\cite{Mapping}
$V \leftrightarrow J_{\perp}/\sqrt{8}$ and
$\delta_{\rm U} \leftrightarrow
\sqrt{2} \delta_z + \pi(\sqrt{2} - 1)/2$, with
$\delta_{\rm U} = -\arctan(\pi \varrho_0 U/2)$
and $\delta_z = -\arctan(\pi \varrho_0 J_z/4)$.
Our NRG results for the low-energy scales of both models
are summarized in Fig.\ref{fig:resonant_vs_kondo}.
Evidently, there is close correspondence between the
two models using the above mapping of parameters,
confirming the predictions of bosonization.
Note that $T_K$ varies by a factor of $30$ in
Fig.\ref{fig:resonant_vs_kondo}. The agreement
does not extend to the two-channel IRLM,
which similarly flows to a strong-coupling Fermi-liquid
fixed point (unlike the non-Fermi-liquid fixed
point of the 2CKM).
The DoS renormalization factor, absent in the 1CKM,
proves essential in this case.

---{\em Conclusions}.
We have critically examined the accuracy of the AY
and bosonization methods away from weak coupling by
considering two generic impurity models. Reduction
of the conduction-electron DoS, accounted for by the
AY approach but absent in bosonization, was shown to
be vital in the case of an unstable strong-coupling
fixed point, yet superfluous in models where a decoupled
entity with local dynamics is formed.
The two methods thus display complementary accuracies
at strong coupling, controlled by the stability of the
strong-coupling fixed point. Accuracy of these powerful
methods can thus be significantly enhanced by
selectively incorporating the DoS renormalization
factor, making them adequate tools for tackling
strong-coupling physics.

---{\em Acknowledgments}.
We are grateful to Natan Andrei for stimulating discussions.
This research was supported in part by Hungarian Grants
OTKA through project T048782 (L.B. and A.Z.), by the
J\'anos Bolyai Foundation and the Alexander von Humboldt
Foundation (L.B), and by the Israel Science Foundation
(A.S.).


\begin{thebibliography}{99}
\bibitem{Mapping}
   Note that Ref.~\onlinecite{Schlottmann82} settled
   with linearizing the phase shifts. It also lacks
   a factor of $1/\sqrt{2}$ in $J_{\perp}$.
\end{thebibliography}
\end{document}